\documentclass[12pt]{article}
\usepackage[left=2.50cm, right=2.50cm, top=2.50cm, bottom=2.50cm]{geometry}
\usepackage[utf8]{inputenc}
\usepackage{cancel}
\usepackage[english]{babel}
\usepackage{physics}
\usepackage{hyperref}
\usepackage{amsthm}
\usepackage{mathtools}
\usepackage{graphicx}
\usepackage{changepage}
\usepackage{amssymb}
\usepackage{verbatim}
\usepackage{empheq}
\usepackage{xcolor}
\usepackage{tikz}
\usepackage{multirow}
\usepackage{multicol}
\usepackage{amsmath}
\usepackage{centernot}
\usepackage[normalem]{ulem}
\usepackage[hang,flushmargin]{footmisc} 
\newcommand{\be}{\begin{equation}}
\newcommand{\ee}{\end{equation}}
\newcommand{\bea}{\begin{eqnarray}}
\newcommand{\eea}{\end{eqnarray}}
\usetikzlibrary{tikzmark}
\numberwithin{equation}{section}
\hypersetup{hidelinks}

\newlength{\bibitemsep}
\newlength{\bibparskip}
\let\oldthebibliography\thebibliography
\renewcommand\thebibliography[1]{%
  \oldthebibliography{#1}%
  \setlength{\parskip}{\bibitemsep}%
  \setlength{\itemsep}{\bibparskip}%
}
\begin{document}
\par
\bigskip
\Large
\noindent
{\bf Topological BF description of 2D accelerated chiral edge modes
}
\bigskip
\par
\rm
\normalsize

\hrule

\vspace{1cm}

\large
\noindent
{\bf Erica Bertolini$^{1,2,a}$}, {\bf Filippo Fecit$^{1,2,b}$} and
{\bf Nicola Maggiore$^{1,2,c}$}\\

\par
\small
\noindent$^1$ Dipartimento di Fisica, Universit\`a di Genova, Italy.
\smallskip

\noindent$^2$ Istituto Nazionale di Fisica Nucleare - Sezione di Genova, Italy.

\smallskip

\vspace{.5cm}

\noindent
{\tt Abstract~:}
In this paper we consider the topological abelian BF theory with radial boundary on a generic 3D manifold. Our aim is to study if, where and how the boundary keeps memory of the details of the background metric. We find that some features are topologically protected and do not depend on the bulk metric. These are the Ka\c{c}-Moody algebra formed by the conserved currents and its central charge, which is proportional to the inverse of the bulk action coupling constant. We then derive the 2D action holographically induced on the boundary, which depends on two scalar fields, and which can be decoupled in two Luttinger actions describing two chiral bosons moving on the edge of the 3D bulk. The outcome is that these edge excitations are $accelerated$, as a direct consequence of the non-flat nature of the bulk spacetime. The chiral velocities of the edge modes, indeed, acquire a local dependence through the determinant of the induced metric on the boundary. 
We find three possibilities for the motion of the edge quasiparticles: same directions, opposite directions and a single-moving mode. But, requiring that the Hamiltonian of the 2D theory is bounded by below, {\it the case of edge modes moving in the same direction is ruled out}: systems involving parallel Hall currents (for instance Fractional Quantum Hall Effect with $\nu=2/5$)  cannot be described by a BF theory with boundary, independently from the geometry of the bulk spacetime, because of positive energy considerations. We are therefore left with physical situations characterized by edge excitations moving with opposite velocities (examples are FQHE with $\nu=1-1/n$, with $n$ positive integer, and Helical Luttinger Liquids phenomena) or a single-moving mode (Quantum Anomalous Hall). A strong restriction is obtained by requiring Time Reversal symmetry, which uniquely identifies modes with equal and opposite  velocities, and we know that this is the case of Topological Insulators. The novelty, with respect to the flat bulk background, is that the modes have local velocities, which corresponds to Topological Insulators with $accelerated$ edge modes.

\vspace{.5cm}

\vspace{\fill}
\noindent{\tt Keywords:} 
Quantum Field Theory, Field Theory with Boundary, Topological States of Matter.

\vspace{.5cm}

\hrule
\noindent{\tt 
$^a$erica.bertolini@ge.infn.it,
$^b$filippo.fecit@ge.infn.it,
$^c$nicola.maggiore@ge.infn.it.
}
\newpage
\section{Introduction}

Topological Quantum Field Theories (TQFT) are characterized by the defining property that their observables do not depend on the metric of the background spacetime. As a consequence, TQFT do not display $physical$ observables, which are local, but ``only'' $geometrical$ ones, $i.e.$ global properties of the manifolds, like handles, knots and so on \cite{Witten:1988ze,Birmingham:1991ty}. An easy way to highlight this peculiar property is to compute the energy-momentum tensor of a TQFT:
the result is zero or, more precisely, the only contribution comes from the unphysical gauge-fixing sector. The situation drastically changes if a boundary is introduced in the background spacetime. The presence of a boundary breaks everything which can be broken, starting from translations and rotations, hence Lorentz invariance. In TQFT the boundary also breaks gauge invariance, since both Schwarz-type TQFT Lagrangians (Chern-Simons (CS) in 3D and BF in any spacetime dimensions) transform as total derivatives under gauge transformations  \cite{Birmingham:1991ty}. Hence, if a boundary is present, integration by parts gives a nonvanishing contribution. On the lower dimensional boundary unbroken residual symmetries survive, unless a boundary on the boundary is placed again, which might be interesting \cite{Schindler}. The question naturally arises of which boundary conditions (BC) should be imposed on the bulk fields or, rather, which are the most general BC which naturally emerge as a consequence of the presence of a boundary itself, without introducing them by hand, in order to get rid of the dependence on any particular choice. The other crucial issue is that of determining which are the remnants of the bulk theory on the boundary physics, namely which is the holographic projection of the $d$-dimensional bulk theory on the $(d-1)$-dimensional boundary. Boundary effects in physics are studied since ever. In the framework of QFT, a pioneering work has been done by Symanzik in the case of the Casimir effect \cite{Symanzik:1981wd}, which is the textbook example of the physical consequences of the presence of boundaries. In \cite{Symanzik:1981wd}, Symanzik introduced some of the basic tools for dealing with a boundary in QFT. For instance, in this paper we adopted Symanzik's way of determining the most general BC, thus avoiding the arbitrariness of an imposition by hand. It basically goes as follows. Given that the boundary breaks the breakable, a boundary term should be added to the action, constrained only by the most general requests of field theory: locality and power counting \cite{Blasi:2010gw,Amoretti:2013nv,Amoretti:2014kba,Bertolini:2020hgr}. The BC are then determined by a variational principle, applied on the equations of motion (EOM) of the theory, modified by the presence of the boundary term. The study of TQFT with boundary led to remarkable results, mainly in condensed matter physics. The Hall systems have been understood in terms of a 3D CS theory with planar boundary \cite{Stone:1990iw,Wen:1992vi,Blasi:2008gt,Maggiore:2017vjf}, and the Topological Insulators (TI), which represent the other important topological phase of matter \cite{moorenature,Hasan:2010xy,Hasan:2010hm,Qi-Zhang}, are described by the topological BF theory with planar boundary, in 3D and 4D \cite{Cho:2010rk,Blasi:2011pf,Amoretti:2012hs,Amoretti:2014iza}. In both cases it has ben possible to show the existence, on the boundary, of conserved currents forming an algebra of the Ka\c{c}-Moody (KM) type \cite{Kac:1967jr,Moody:1966gf}, with central charge proportional to the inverse of the action coupling constant. The holographic (in the sense previously explained) 2D theory induced on the boundary of the abelian CS theory is the Floreanini-Jackiw action \cite{Floreanini:1987as}, describing edge modes moving with constant chiral velocities, which are indeed observed \cite{kane87}. The velocity of the edge excitations is the main observable of both Hall systems and TI, and it turns out to be related to the action coupling constant or, equivalently, to the central charge of the boundary KM algebra formed by the conserved currents. Recently, Hall systems edge modes  have been observed displaying $non\ constant$ velocities \cite{Bocquillon}. These accelerated chiral modes cannot be explained by the above description (TQFT on flat spacetime with planar boundary), which unavoidably yields $constant$ edge velocities. The standard approach to deal with these cases is phenomenological, and basically consists in adding a suitable potential to the 2D Luttinger action, in order to reproduce time-dependent velocities \cite{Wen:1989mw,Wen:1990qp,Kane95,Hashi18,Wen:1991ty}. The price of this way of solving the problem is that the whole ``holographic'' construction of finding the right lower dimensional action without any $ad\ hoc$ extension fails. Recently, an alternative approach has been proposed to reproduce accelerated edge modes in Hall systems without the need of adding any empirical potential, while keeping the holographic construction intact \cite{Bertolini:2021iku}. It consists in considering the bulk theory on a generic, rather that flat, background manifold. While the topological invariant action of course does not depend on the particular spacetime metric, Symanzik's boundary term and the boundary itself certainly do, and it is an interesting issue to find out if, how and where this metric dependence reflects on the holographic 2D theory and, more interestingly, on physical observables. Remarkably, it has been shown that, indeed, a remnant of the bulk metric remains on the most relevant physical observable, $i.e.$ the chiral velocity of the edge modes, which becomes $local$. In other words, the edge modes of the Hall systems, when described by a TQFT built on a generic manifold, are accelerated and, moreover, the velocities also depend on the position of the quasiparticle on the boundary, not only on time. The dependence on the bulk metric manifests only through the determinant of the induced metric on the boundary, hence it is mild, as one might expect due to the topological character of the bulk theory. Still, the local effects are reproduced, without any empirical modification of the 2D holographic Luttinger theory. The aim of this paper is to see whether a similar result holds also for TI, when described by a topological abelian 3D BF theory on a generic manifold with radial boundary. This paper is organized as follows. In Section 2 we prepare the tools to face the problem: we introduce a radial boundary in the gauge-fixed 3D BF action, and derive {\it \`a la} Symanzik the most general BC on the two gauge fields of the theory. As we said, the boundary breaks gauge invariance, and this reflects in the breakings of the two Ward identities describing the broken gauge symmetry. The breaking are particularly fruitful, as they lead us to identify the 2D scalar degrees of freedom and the KM algebra formed by the edge conserved currents. Requiring the positivity of the KM central charge constrains the BF coupling to be positive. In Section 3 the 2D theory is derived as the holographic projection of the 3D bulk theory. The contact is realized by matching the BC on the bulk side and the EOM on the boundary side. The resulting action involves two scalar fields and is more complicated than the simple Luttinger theory found on the edge of CS model. We give a physical interpretation of the 2D theory, and we find three possibilities for the motion of the edge quasiparticles: same directions, opposite directions and a single-moving mode. But, requiring that the Hamiltonian of the 2D theory is bounded by below, {\it the case of edge modes moving in the same direction is ruled out}: systems involving parallel Hall currents (for instance Fractional Quantum Hall Effect with $\nu=2/5$ \cite{wen2})  cannot be described by a BF theory with boundary, independently from the geometry of the bulk spacetime, just because of positive energy considerations. We are therefore left with physical situations characterized by edge excitations moving with opposite velocities (examples are FQHE with $\nu=1-1/n$, with $n$ positive integer \cite{wen2}, and Helical Luttinger Liquids phenomena, characterized by the existence of two conserved chiral edge currents propagating in opposite directions with same speed and definite spin \cite{Wu2006HelicalLA}) or a single-moving mode (Quantum Anomalous Hall \cite{Qi-Zhang, Liu, Yu}). In Section 4 a strong restriction is obtained by requiring Time Reversal symmetry, which uniquely identifies modes with equal and opposite  velocities, and we know that this is the case of Topological Insulators. The novelty, with respect to the flat bulk background, is that the modes have local velocities, which corresponds to Topological Insulators with $accelerated$ edge modes. In Section 5 we summarize and discuss our results.

\subsection{Notations and conventions}

Indices and coordinates:
\begin{equation}
	\begin{split}
	\mu,\nu,\rho,...=&\{0,1,2\}=\{t,r,\theta\}\\
	i,j,...=& \{0,2\}=\{t,\theta\}\ ,
	\end{split}\label{1.1}
\end{equation}
\begin{equation}
\begin{split}
x=&(x_0,x_1,x_2)=(t,r,\theta)\label{1.2}\\
X=&(x_0,x_2)=(t,\theta)\ .
\end{split}
\end{equation}
The Levi-Civita tensor $\epsilon^{\mu\nu\rho}$ is written in terms of the corresponding symbol $\tilde\epsilon^{\mu\nu\rho}$ as follows
\begin{equation}
\epsilon^{\mu\nu\rho}=\frac{\tilde\epsilon^{\mu\nu\rho}}{\sqrt{-g}}\ ,
\label{}\end{equation}
where $g$ is the determinant of the bulk metric tensor $g_{\mu\nu}$, with Lorentzian signature. The scalar Dirac delta distribution $\delta^{(n)}(x-x')$ and the {corresponding} density $\tilde\delta^{(n)}(x-x')$ are related by \cite{Basler:1991st}
\begin{equation}
\delta^{(n)}(x-x') = \frac{\tilde\delta^{(n)}(x-x')}{\sqrt{-g}}\ ,
\label{1.5}\end{equation}
acting on a test function $f(x)$ as
\begin{equation}
\int d^nx\,\sqrt{-g}\,\delta^{(n)}(x-x')f(x) = 
\int d^nx\,\tilde\delta^{(n)}(x-x')f(x)
=f(x')\ .
\label{}\end{equation}
The functional derivative is defined as
\begin{equation}
\frac{\delta V_\mu(x)}{\delta V_\nu(x')}=\delta^\nu_\mu\,\delta^{(3)}(x-x')\ .
\label{fd}\end{equation}
For a generic manifold $\mathcal{V}$, Stokes' theorem states that
\begin{equation}\label{1.6}
\int_\mathcal{V}d^nx\; \sqrt{-g}\, \nabla_\mu V^\mu 
=
\int_{\partial \mathcal{V}} d^{n-1}y\; \sqrt{-\gamma}\; e_\mu V^\mu\ ,
\end{equation}
where $e_\mu$ is the unit vector normal to the boundary $\partial\mathcal V$ described by the equation $f(x)=0$, $i.e.$
	\begin{equation}
	e_\mu=-\frac{\partial_\mu f}{\sqrt{g^{\mu\nu}\partial_\mu f\partial_\nu f}}\ ,
	\end{equation}
and $\gamma_{ij}$ is the induced metric on $\partial\mathcal V$.
The derivative of the Heaviside step-function is \cite{Bertolini:2021iku}
	\begin{equation}\label{1.9x}
	\nabla_\mu\theta(f(x))=\partial_\mu\theta\left(f(x)\right)=-e_\mu\;\tfrac{\sqrt{|\gamma|}}{\sqrt{|g|}}\delta(f(x))\ .
	\end{equation}
In this paper we will consider a constant, radial boundary
	\begin{equation}\label{n}
	f(x)=R-r\quad\Rightarrow\quad e_\mu=\frac{\delta_\mu^r}{\sqrt{g^{rr}}}=\delta_\mu^r\frac{\sqrt{|g|}}{\sqrt{|\gamma|}}\ ,
	\end{equation}
where we used (eq.(16.42) of \cite{Blau})~:
	\begin{equation}\label{medamath}
	\sqrt{g^{rr}}=\frac{\sqrt{-\gamma}}{\sqrt{-g}}\ .
	\end{equation}
Therefore, using \eqref{n}, the derivative of the step function \eqref{1.9x} simplifies to
	\begin{equation}\label{1.9}
	\nabla_\mu\theta(R-r)
	=\partial_\mu\theta\left(f(x)\right)=-\delta_\mu^r\;\delta(r-R)\ .
	\end{equation}

\section{The model: bulk and boundary}

\subsection{The action}

We consider the abelian BF model on a manifold diffeomorphic to a cylinder of radius $R$. The boundary is introduced by means of a Heaviside step function in the bulk action, constraining the radial coordinate to $r\leq R$. The BF bulk action is 
\begin{equation}\label{2.1}
	S_{BF}=\kappa\int d^3x\,\theta(R-r)\,\tilde\epsilon^{\mu\nu\rho}\partial_\mu A_\nu B_{\rho}\ ,
	\end{equation}
where $A_\mu$ and $B_\mu$ are two gauge fields with mass dimensions $[A]=[B]=1$ and $\kappa$ is a constant which will be determined by physical inputs, as we shall see later. We choose the radial gauge, implemented by the following gauge-fixing term
	\begin{equation}
	S_{gf}=\int d^3x\sqrt{-g}\,\theta(R-r)\left[\left(b A_\mu+d\; B_{\mu}\right) n^\mu\right]\ ,
	\end{equation}
where $n^\mu=(0,1,0)$ is a vector and $b,d$ are scalar Nakanishi-Lautrup Lagrange multipliers \cite{Nakanishi:1966zz,Lautrup:1967zz}:
	\begin{equation}
	\frac{\delta S}{\delta b}=n^\mu A_\mu=A_r=0\quad;\quad	\frac{\delta S}{\delta d}=n^\mu B_\mu=B_r=0\ .
	\end{equation}
The external source term is
	\begin{equation}
	S_{ext}=\int d^3x\sqrt{-g}\,\theta(R-r)\,\left(J^\mu A_\mu+\hat J^\mu B_\mu\right)\ ,
	\end{equation}
where $J^\mu$ and $\hat J^\mu$ are vectors. The presence of a boundary induces, as an additional contribution, the most general boundary term compatible with power-counting and locality \cite{Symanzik:1981wd}
	\begin{equation}\label{2.5}
	S_{bd}=\int d^3x\,\sqrt{-\gamma}\,\delta(r-R)\left(\frac{\alpha^{ij}}{2}A_i A_j+\frac{\beta^{ij}}{2}B_iB_j+\zeta^{ij}A_i B_j\right)\ ,
	\end{equation}
where $\alpha^{ij}=\alpha^{ji},\  \beta^{ij}=\beta^{ji}$ and $\zeta^{ij}$ are dimensionless tensors which depend, at most, on the induced metric (components $\gamma^{ij}$ and/or determinant $\gamma$) in the following way
	\begin{eqnarray}
	\alpha^{ij}=&\hat\alpha^{ij}({\gamma})+\hat\alpha({\gamma})\gamma^{ij}&=\hat\alpha^{ij}+\hat\alpha\gamma^{ij}\label{coeff1}\\
	\beta^{ij}=&\hat\beta^{ij}({\gamma})+\hat\beta({\gamma})\gamma^{ij}&=\hat\beta^{ij}+\hat\beta\gamma^{ij}\label{coeff2}\\
	\zeta^{ij}=&\hat\zeta^{ij}({\gamma})+\hat\zeta({\gamma})\gamma^{ij}&=\hat\zeta^{ij}+\hat\zeta\gamma^{ij}\ .\label{coeff3}
	\end{eqnarray}
In the flat limit the parameters $\alpha^{ij}$, $\beta^{ij}$ and $\zeta^{ij}$ in $S_{bd}$ are constant \cite{Blasi:2019wpq}. Finally, the total action, containing BF bulk, gauge fixing, external sources and boundary terms, is
	\begin{equation}\label{2.8}
	S=S_{BF}+S_{gf}+S_{ext}+S_{bd}\ .
	\end{equation}
	
\subsection{Equations of motion and boundary conditions}\label{secBC}

From the total action $S$ \eqref{2.8}, the EOM of the theory follow
	\begin{equation}
	\frac{\delta S}{\delta A_\lambda}=\theta(R-r)\left[-\kappa\epsilon^{\mu\lambda\rho}\partial_\mu B_\rho+b\, n^\lambda+J^\lambda\right]+\delta^\lambda_i\frac{\sqrt{-\gamma}}{\sqrt{-g}}\delta(r-R)\left[-\kappa\frac{\tilde\epsilon^{i1j}}{\sqrt{-\gamma}}B_j+\alpha^{ij}A_j+\,\zeta^{ij}B_j\right]=0\label{2.9}
	\end{equation}
	\begin{equation}
	\frac{\delta S}{\delta B_\lambda}=\theta(R-r)\left[\kappa\epsilon^{\mu\nu\lambda}\partial_\mu A_\nu+d\, n^\lambda+\hat J^{\lambda}\right]+\delta^\lambda_i\frac{\sqrt{-\gamma}}{\sqrt{-g}}\delta(r-R)\left[\beta^{ij}B_j+\zeta^{ji}A_j\right]=0\ .\label{2.12}
	\end{equation}
From the EOM \eqref{2.9}, \eqref{2.12} we get the boundary conditions (BC) of the theory by applying the operator $\lim_{\epsilon\to R}\int^R_\epsilon$, $i.e.$\\
	\begin{empheq}{align}
	\lim_{\epsilon\to R}\int^R_\epsilon dr\eqref{2.9} :\qquad&\left.\alpha^{ij}A_j+\left(-\kappa\frac{\tilde\epsilon^{i1j}}{\sqrt{-\gamma}}+\,\zeta^{ij}\right)B_j\right|_{r=R}=0\label{BC1}\\
	\lim_{\epsilon\to R}\int^R_\epsilon dr\eqref{2.12} :\qquad&\left.\zeta^{ji}A_j+\beta^{ij}B_j\right|_{ r=R}=0\ .\label{BC2}
	\end{empheq}
Eq.\eqref{BC1} and \eqref{BC2} can be written in a compact form as follows
	\begin{equation}\label{B.16}
	\left.\Lambda^{IJ}X_J\right|_{r=R}=0\ ,
	\end{equation}
where $I,J=\{i;j\}=\{0,2;0,2\}$,
\begin{equation}
\Lambda^{IJ}\equiv\left(\begin{array}{cc}
\alpha^{ij} & \zeta^{i j}-\kappa\epsilon^{i1 j} \\
\zeta^{j i} & \beta^{ i j} \\
\end{array}\right)=\left(
\begin{array}{cccc}
\alpha^{00} & \alpha^{02} & \zeta^{00} & \zeta^{02}-\hat\kappa\\
 \alpha^{20} & \alpha^{22} & \zeta^{20}+\hat\kappa & \zeta^{22}\\
 \zeta^{00} & \zeta^{20} & \beta^{00} & \beta^{02}\\
 \zeta^{02} & \zeta^{22}& \beta^{02} & \beta^{22}
\end{array}
\right)\ , \label{Lambda}
\end{equation}
and
\begin{equation}
X_J\equiv\left(\begin{array}{c}
A_j\\
B_{j}
\end{array}\right)\ ,
\label{B.18}
\end{equation}
where we defined
\begin{equation}\label{B.19}
\hat\kappa\equiv\frac{\kappa\tilde\epsilon^{012}}{\sqrt{-\gamma}}\ .
\end{equation}
We leave $\tilde\epsilon^{012}$ explicit (instead of simply putting it equal to 1) to keep explicit the tensor nature of all quantities. For instance, in this way, it is immediate to see that $\hat\kappa$ is a scalar function. The BC \eqref{B.16} defines a linear, homogeneous system of four equations and four variables $A_i|_{r=R}$ and $B_i|_{r=R}$, for which, requiring $\det\Lambda=0$, it is possible to write three of the fields in terms of one:
	\begin{empheq}{align}
	B_\theta(X)&= -l_1B_t(X)\label{bcs2'}\\
	A_\theta(X)&= -l_2B_t(X)\label{bcs4'}\\
	A_t(X)&= -l_3B_t(X)\label{bcs6'}\ ,
	\end{empheq}
where $l_{1,2,3}$ depend on the coefficients of the total action \eqref{2.8} and therefore, in general, are local functions of the induced metric on the boundary $\gamma_{ij}(X)$. Their explicit form can be found in Appendix \ref{appl} (\eqref{l1}, \eqref{l2} and \eqref{l3}). Notice that to exclude Dirichelet-like solutions ($i.e.\ A_i|_{r=R}=B_i|_{r=R}=0$), which would trivialize the boundary 2D physics,  we must require  $l_i\neq0$ and $l_i^{-1}\neq0$.

\subsection{Ward identities}

The covariant divergence of the EOM \eqref{2.9} is
\begin{equation}\label{2.17}
		\begin{split}
		\nabla_\lambda\frac{\delta S}{\delta A_\lambda}=&\nabla_\lambda\left[\theta(R-r)\left(-\kappa\epsilon^{\mu\lambda\rho}\partial_\mu B_\rho+ n^\lambda\,b+J^\lambda\right)\right]\\
		=&\frac{1}{\sqrt{-g}}\delta(r-R)\,\kappa\,\tilde\epsilon^{i1j}\partial_i B_j+\frac{1}{\sqrt{-g}}\partial_\lambda\left[\theta(R-r) n^\lambda\,b\sqrt{-g}\right]+\nabla_\lambda\left[\theta(R-r)J^\lambda\right]=0\ ,
		\end{split}
	\end{equation}
where we used the BC \eqref{BC1}, the fact that  $\epsilon^{\mu\nu\rho}\nabla_\mu\partial_\nu B_{\rho}=0$, the definition of covariant derivative of the step function \eqref{1.9}  and the formula for the covariant divergence
	\begin{equation}\label{div-cov'}
	\nabla_\mu V^\mu=\frac{1}{\sqrt{-g}}\partial_\mu\left(V^\mu\sqrt{-g}\right)\ .
	\end{equation} 
By multiplying \eqref{2.17} by $\sqrt{-g}$ and integrating over the coordinate normal to the boundary $r=R$, we get
	\begin{equation}\label{int ward}
		\begin{split}
		&\int_0^{+\infty}dr\,\left\{\delta(r-R)\,\kappa\,\tilde\epsilon^{i1j}\partial_i B_j+\partial_\lambda\left[\theta(R-r) n^\lambda\,b\sqrt{-g}\right]+\sqrt{-g}\,\nabla_\lambda\left[\theta(R-r)J^\lambda\right]\right\}\\
		=&\left.\kappa\tilde\epsilon^{i1j}\partial_iB_j\right|_{r=R}-\cancel{b\,\sqrt{-g}|_{r=0}}+\int_0^{+\infty}dr\,\sqrt{-g}\,\nabla_\lambda\left[\theta(R-r)J^\lambda\right]=0\ ,
		\end{split}
	\end{equation}
where we adopted the convention according to which the $\theta$ components of the fields, their $\theta$-derivatives and the Lagrange multipliers vanish at $r=0$ :
	\begin{equation}
	\left.A_\theta=B_\theta=\partial_\theta A_t=\partial_\theta B_t=b=d\right|_{r=0}=0\ .
	\end{equation}
A comment here may be useful: the invariant measure for integrating along $r$ can be identified as induced from the bulk, as follows
	\begin{equation}
	\int d^3x\sqrt{-g}\delta^{(2)}(X-X')=\int d^3x\sqrt{-g}\frac{\tilde\delta^{(2)}(X-X')}{\sqrt{-\gamma}}=\int dr\frac{\sqrt{-g}}{\sqrt{-\gamma}}=\int dr\,\sqrt{g_{rr}}\ ,
	\end{equation}
 where we used \eqref{medamath} and $X=(t,\theta)$ are the boundary coordinates. However, in the specific case of \eqref{int ward}, the $\sqrt{g_{rr}}$ factor in the integration can be omitted, simply dividing \eqref{2.17} by $\sqrt{g_{rr}}$. Using \eqref{1.9} in \eqref{int ward}, we finally find
	\begin{equation}\label{ward1}
	\int_0^Rdr\,\sqrt{-g}\nabla_\lambda J^\lambda=\left(-\kappa\tilde\epsilon^{i1j}\partial_iB_j+\sqrt{-g}J^r\right)_{r=R}\ ,
	\end{equation}
which is the Ward identity corresponding to the gauge transformation of the gauge field $A_\mu$, broken at its rhs by the presence of the boundary. By applying the same procedure to the EOM \eqref{2.12}, with $d|_{r=0}=0$, we obtain a second broken Ward identity:
	\begin{equation}\label{ward2}
	\int_0^Rdr\,\sqrt{-g}\nabla_\lambda \hat J^\lambda=\left(-\kappa\tilde\epsilon^{i1j}\partial_iA_j+\sqrt{-g}\hat J^r\right)_{r=R}\ .
	\end{equation}
From \eqref{ward1} and \eqref{ward2}, going on-shell ($i.e.\  J=\hat J=0$), we find
	\begin{empheq}{align}
	\left.\tilde\epsilon^{i1j}\partial_i B_{j}\right|_{r=R}=&0\label{2.26}\\
	\left.\tilde\epsilon^{i1j}\partial_i A_j\right|_{r=R}=&0\ ,\label{2.27}
	\end{empheq}
which describe two conserved currents on the boundary $r=R$. The most general solutions to these equations are \cite{Nash:1983cq,Warner}
	\begin{empheq}{align}
	B_i(X)=&\partial_i\psi(X)+\delta_{i2}\hat C\label{2.28}\\
	A_i(X)=&\partial_i\varphi(X)+\delta_{i2}C\ ,\label{2.29}
	\end{empheq}
where $C$ and $\hat C$ are two constants and $\varphi(X)$ and $\psi(X)$ are scalar boundary fields  with zero mass dimensions, $i.e.\ [\varphi]=[\psi]=0$, which should be identified with the boundary degrees of freedom (DOF). Being on a closed, periodic boundary, we need to specify periodicity conditions
	\begin{empheq}{align}
	A_i(t,\theta)=A_i(t,\theta+2\pi ) &\Rightarrow \varphi(t,\theta)=\varphi(t,\theta+2\pi )\label{period1'}\\
	B_i(t,\theta)=B_i(t,\theta+2\pi ) &\Rightarrow \psi(t,\theta)=\psi(t,\theta+2\pi )\ .\label{period2'}
	\end{empheq}
The values of the constants $C$ and $\hat C$ in \eqref{2.28}, \eqref{2.29} are found by applying the mean value theorem for holomorphic functions, which states that if $f$ is analytic in a region $D$, and $a\in D$, then $f(a)=\frac{1}{2\pi}\oint_{{\cal C}(a)} f$, where ${\cal C}(a)$ is a circle centered in $a$. In our case (2+1 dimensions) taking  for $\cal C$ the circular boundary $r=R$ centered at $r=0$ allows us to write
	\begin{equation}
	A_{\theta}(t,r=0,\theta)=\oint_{ring\; R} A_\theta(x) 
	=\cancel{\left.\varphi(t,\theta)\right|_{\theta=0}^{\theta=2\pi }} +
	 2\pi \; C\ .
	\label{holo A}
	\end{equation}
From the requirement $A_{\theta}(t,r=0,\theta)=0$, it follows that
	\begin{equation}
	C=0\ .
	\label{C=0}
	\end{equation}
Analogously, we also get
	\begin{equation}
	\hat C=0\ .
	\label{hatC=0}
	\end{equation}

\subsection{Algebra}

The generating functional of the connected Green functions $Z_c[J,\hat J]$ is defined in the usual way
\begin{equation}
e^{iZ_c[J,\hat J]}=\int DA\;DB\;Db\;Dd\;e^{iS[A,B,b,d;J,\hat J]}\  ,
\label{xA.36}
\end{equation}
from which the 1- and 2-points Green functions are derived
\begin{empheq}{align}
\left.\frac{\delta Z_c[J]}{\delta J^i(x)}\right|_{J=0}=&\;\langle A_i(x)\rangle\label{xA.37}\\
\left.\frac{\delta^{(2)} Z_c[J]}{\delta J^i(x)\delta J^j(x')}\right|_{J=0}\equiv&\Delta_{ij}(x,x')=i\langle T(A_i(x)A_j(x'))\rangle\ ,\label{xA.38}
\end{empheq}
where $T$ is the time-ordered product
\begin{equation}
\langle T(A_l(x)A_j(x'))\rangle\equiv\theta(t-t')\langle A_l(x)A_j(x') \rangle +
\theta(t'-t)\langle A_j(x') A_l(x) \rangle\ .
\label{xA.39}
\end{equation}
In order to compute the propagator \eqref{xA.38}, we need the following result
	\begin{equation}\label{funcJ}
		\frac{\delta}{\delta J^i(x')}\nabla_\lambda J^\lambda(x)=\frac{\delta}{\delta J^i(x')}\left[\frac{1}{\sqrt{-g}}\partial_\lambda\left(J^\lambda\sqrt{-g}\right)\right]
		=\frac{1}{\sqrt{-g}}\partial_i\tilde\delta^{(3)}(x-x')\ ,
	\end{equation}
where we used \eqref{div-cov'} and the relation between scalar and density Dirac delta distributions \eqref{1.5}. Taking now the functional derivative with respect to the external sources $J,\hat J$ of the  broken Ward identities \eqref{ward1} and \eqref{ward2}, we get\\[10px]
$\frac{\delta}{\delta J^k(x')}\eqref{ward1}\;:$
	\begin{equation}\label{2.30}
		\begin{split}
		\int^R_0dr \;\partial_k\tilde\delta^{(3)}(x-x')&=\left.- \kappa\tilde\epsilon^{i1j}\partial_i\frac{\delta^{(2)}Z_c}{\delta J^k(X')\delta\hat J^{j}(X)}\right|_{J=\hat J=0}\\
		\partial_k\tilde\delta^{(2)}(X-X')&=-i\kappa\tilde\epsilon^{i1j}\partial_i\langle T\left(A_k(X')B_{j}(X)\right)\rangle\\
		&=-i\kappa\tilde\epsilon^{012}\left[B_{\theta}(X),A_k(X')\right]\partial_t\theta(t-t')\ ,
		\end{split}
	\end{equation}
where we used \eqref{xA.38},  \eqref{xA.39} and \eqref{funcJ}. To write \eqref{2.30} we used the fact that for any $r'\leq R$ we have $\int_0^Rdr\,\tilde\delta(r-r')=1$, in fact, since by definition $\tilde\delta(r-r')\equiv-\partial_r\theta(r'-r)$:
	\begin{equation}
		\begin{split}
		\int_0^Rdr\,\partial_r\theta(r'-r)f(r)&=\int_0^R\left\{\partial_r\left[\theta(r'-r)f(r)\right]-\theta(r'-r)\partial_rf(r)\right\}\\
		&=\left.\theta(r'-r)f(r)\right|_0^R-\int_0^Rdr\,\theta(r'-r)\partial_rf(r)\\
		&=	\left\{\begin{array}{cccccc}
			f(R)-f(0)-\int^R_0dr\,\partial_rf(r)&=&0&\ & \mbox{if}\ r'>R&\\
			0-f(0)-\int^{r'}_0dr\,\partial_rf(r)&=&-f(r')&\ & \mbox{if}\ r'\leq R&\ .
			\end{array}\right.
		\end{split}
	\end{equation}
Going on-shell, $i.e.$ using \eqref{2.26}, we have
	\begin{equation}\label{2.31}
	\tilde\epsilon^{012}\left[B_{\theta}(X),A_k(X')\right]\partial_t\theta(t-t')=\frac{i}{\kappa}\partial_k\tilde\delta^{(2)}(X-X')\ .
	\end{equation}
Setting $k=t$ in \eqref{2.31} and integrating over time, we get the equal time commutator
	\begin{equation}\label{2.32}
	\left[ B_\theta(X),A_t(X')\right]=0\ .
	\end{equation}
Analogously, choosing $k=\theta$ we get 
	\begin{equation}\label{3.41}
	\left.\tilde\epsilon^{012}\left[B_\theta(X),A_\theta(X')\right]\right|_{t=t'}=\frac{i}{\kappa}\partial_\theta\tilde\delta(\theta-\theta')\ .
	\end{equation}
We can repeat this to find the whole current algebra\\
$\frac{\delta}{\delta \hat J^{k}(x')}\eqref{ward1}\;:$\\
	\begin{equation}\label{2.34}
		0=\left.\tilde\epsilon^{i1j}\partial_i\frac{\delta^{(2)}Z_c}{\delta\hat J^{k}(X')\delta\hat J^{j}(X)}\right|_{J=0}=\tilde\epsilon^{i1j}\partial_i\langle T\left(B_{k}(X')B_{j}(X)\right)\rangle=\tilde\epsilon^{012}\left[B_{\theta}(X),B_{k}(X')\right]\partial_t\theta(t-t')\ ,
	\end{equation}
which leads to
	\begin{equation}\label{2.35}
	\left.\left[ B_\theta(X),B_{k}(X')\right]\right|_{t=t'}=0\ .
	\end{equation}
In the same way, from $\frac{\delta}{\delta J^k(x')}\eqref{ward2}$ we get
	\begin{equation}\label{2.37}
	\left[A_\theta(X),A_k(X')\right]_{t=t'}=0\ .
	\end{equation}
Finally, from $\frac{\delta}{\delta\hat J^{k}(x')}\eqref{ward2}$ we have
	\begin{empheq}{align}
	\left[A_\theta(X),B_{t}(X')\right]_{t=t'}=&0\\
	\tilde\epsilon^{012}\left[A_\theta(X),B_{\theta}(X')\right]_{t=t'}=&\frac{i}{\kappa}\partial_\theta\tilde\delta(\theta-\theta')\ .\label{3.55}
	\end{empheq}
Summarizing, the equal time commutators are
	\begin{empheq}{align}
	\left[ B_\theta(X),A_t(X')\right]=&0\label{2.41}\\
	\tilde\epsilon^{012}\left[B_\theta(X),A_\theta(X')\right]=&\frac{i}{\kappa}\partial_\theta\tilde\delta(\theta-\theta')\label{2.42}\\
	\left[ B_\theta(X),B_{k}(X')\right]=&0\label{2.43}\\
	\left[A_\theta(X),A_k(X')\right]=&0\label{2.44}\\
	\left[A_\theta(X),B_{t}(X')\right]=&0\label{2.45}\\
	\tilde\epsilon^{012}\left[A_\theta(X),B_{\theta}(X')\right]=&\frac{i}{\kappa}\partial_\theta\tilde\delta(\theta-\theta')\ .\label{2.46}
	\end{empheq}
One can observe that by using the property of the delta function $\delta'(x)=-\delta'(-x)$, the commutators \eqref{2.42} and \eqref{2.46} represent the same relation. 
 Eq.\eqref{2.41}-\eqref{2.46} describe  a semidirect sum of Kaç-Moody (KM) algebras \cite{Kac:1967jr,Moody:1966gf} with central charge
	\begin{equation}\label{2.66}
	c=\frac{1}{\kappa}\ ,
	\end{equation}
which, as one can expect, inherit the topological nature of the bulk theory, being independent from the metric. It is interesting to remark that from the positivity of the central charge, being necessary for the unitarity of the CFT \cite{mack,Becchi:1988nh}, we get a constraint on the coupling constant of the BF bulk model
	\begin{equation}\label{2.67}
	{\kappa}>0\ .
	\end{equation}
Typically, the constraint on the coupling constants of QFT is derived by asking the positivity of the energy density, which is the 00-component of the energy momentum tensor $T^{00}$. This cannot be achieved in TQFT, which have vanishing Hamiltonian, as it is well known. The constraint in that case is obtained by asking the positivity of the central charge of the edge current algebra, as in the present case. 

\section{2D boundary theory}\label{sec2Dth}

We now focus on the construction of the boundary theory, whose DOF are  the boundary scalar fields $\varphi(X)$ and $\psi(X)$ defined by 
the solutions \eqref{2.28} and \eqref{2.29} of the conserved currents equations \eqref{2.26} and \eqref{2.27}. To build the 2D induced theory we follow three steps:
	\begin{enumerate}
	\item identification of the 2D canonical variables in terms of boundary fields ;
	\item derivation of the most general 2D action ;
	\item  bulk-boundary correspondence (holographic contact).
	\end{enumerate}
	
\subsection{2D canonical variables}\label{sec3.1}

The first step is to write the commutator \eqref{2.46} in terms of the boundary fields $\varphi(X),\psi(X)$, using \eqref{2.28} and \eqref{2.29} (with $C=\hat C=0$) :
	\begin{equation}\label{A.61}
	\tilde\epsilon^{012}\left[\partial_\theta\varphi(X),\partial_{\theta'}\psi(X')\right]=\frac{i}{\kappa}\partial_\theta\tilde\delta(\theta-\theta')\quad\Rightarrow\quad\tilde\epsilon^{012}\left[\varphi(X),\partial_{\theta'}\psi(X')\right]=\frac{i}{\kappa}\tilde\delta(\theta-\theta')\ .
	\end{equation}
Analogously, from \eqref{2.42} we get
	\begin{equation}\label{comm2}
	\tilde\epsilon^{012}\left[\psi(X),\partial_{\theta'}\varphi(X')\right]=\frac{i}{\kappa}\tilde\delta(\theta-\theta')\ .
	\end{equation}
Both the relations \eqref{A.61} and \eqref{comm2} can be interpreted as canonical commutation relations of the type
	\begin{equation}\label{}
	\left[q(X),p(X')\right]=i\tilde\delta(\theta-\theta')\ ,
	\end{equation}
once the following identifications are done:
	\begin{empheq}{align}
	q_1\equiv\varphi\quad;&\quad p_1\equiv\kappa\tilde\epsilon^{012}\partial_\theta\psi\label{A.64}\\
	\color{black}q_2\equiv\psi\quad;&\quad\color{black}p_2\equiv\kappa\tilde\epsilon^{012}\partial_\theta\varphi\ .\normalcolor\label{CV2}
	\end{empheq}
Therefore the DOF of the 2D theory are equivalently described by either of the sets of canonical variables \eqref{A.64} or \eqref{CV2}.

\subsection{The 2D action}\label{sec2Daction}

To find the most general 2D action, we make a derivative expansion in the boundary fields $\varphi(X)$ and $\psi(X)$, compatible with power-counting (we remind that $[\varphi]=[\psi]=0$)
	\begin{equation}\label{A.65}
	S_{2D}[\varphi,\psi]=\int d^2X\,\mathcal L_{2D}=\int d^2X\,\sqrt{-\gamma}\left(a^{ij}\partial_i\varphi\partial_j\varphi+b^{ij}\partial_i\psi\partial_j\psi+c^{ij}\partial_i\varphi\partial_j\psi+d^i\partial_i\varphi+f^i\partial_i\psi+h\right)\ ,
	\end{equation}
where $a^{ij}=a^{ji},$ $b^{ij}=b^{ji}$ and $c^{ij}$ are tensors, $d^i$ and $f^i$ are vectors and $h$ is a scalar, with mass dimensions
	\begin{equation}\label{m dim}
	[a]=[b]=[c]=0\quad;\quad[d]=[f]=1\quad;\quad[h]=2\ .
	\end{equation}
The coefficients appearing in \eqref{A.65} may depend on the induced metric $\gamma_{ij}(X)$ and/or on the boundary fields, but not on their derivatives. The definitions of the scalar fields $\varphi(X)$ and $\psi(X)$ \eqref{2.28} and \eqref{2.29} are invariant under the shift transformations $\delta_s,\delta'_s$ defined as
	\begin{empheq}{align}
	\delta_{s}\varphi=&\eta\label{sh-phi}\\
	\delta'_{s}\psi=&\eta'\ .\label{sh-vphi}
	\end{empheq}
Consequently, the 2D lagrangian $\mathcal L_{2D}$ in \eqref{A.65} must be shift-invariant as well
	\begin{equation}\label{sh-S}
	\delta_s\mathcal L_{2D}=\delta'_s\mathcal L_{2D}=0\ ,
	\end{equation}
which implies
	\begin{equation}\label{A.72}
	S_{2D}[\varphi,\psi]=\int d^2X\,\mathcal{L}_{2D}=\int d^2X\,\sqrt{-\gamma}\left(a^{ij}\partial_i\varphi\partial_j\varphi+b^{ij}\partial_i\psi\partial_j\psi+c^{ij}\partial_i\varphi\partial_j\psi+
d^i\partial_i\varphi+f^i\partial_i\psi\right)\ ,
	\end{equation}
with
\begin{equation}
	\begin{split}
\frac{\partial a^{ij}}{\partial\varphi}=\frac{\partial b^{ij}}{\partial\varphi}=\frac{\partial c^{ij}}{\partial\varphi}=0&\\
\frac{\partial a^{ij}}{\partial\psi}=\frac{\partial b^{ij}}{\partial\psi}=\frac{\partial c^{ij}}{\partial\psi}=0&\ ,
	\end{split}
\end{equation}
and 
\begin{equation}\label{d,f}
	\begin{split}
&\frac{\partial d^{i}}{\partial\varphi}\partial_i\varphi+\frac{\partial f^{i}}{\partial\varphi}\partial_i\psi=0\\
&\frac{\partial d^{i}}{\partial\psi}\partial_i\varphi+\frac{\partial f^{i}}{\partial\psi}\partial_i\psi=0\ .
	\end{split}
\end{equation}
Hence, the coefficients $d^i$ and $f^i$ may still depend on the boundary fields $\varphi(X)$ and $\psi(X)$, provided that the constraints \eqref{d,f} hold.
    In $S_{2D}[\varphi,\psi]$ \eqref{A.72} we omitted the scalar term $h$ because, being metric-dependent only, it does not contribute to the EOM of the boundary theory.
As we did for $S_{bd}$ \eqref{2.5}, we parametrize the metric dependence of the rank-2 tensors as follows:
	\begin{empheq}{align}
	a^{ij}&=\hat a^{ij}+\hat a\,\gamma^{ij}\label{coeff-a}\\
	b^{ij}&=\hat b^{ij}+\hat b\,\gamma^{ij}\label{coeff-b}\\
	c^{ij}&=\hat c^{ij}+\hat c\,\gamma^{ij}\ ,\label{coeff-c}
	\end{empheq}
where the hat means dependence on the metric determinant at most. As observed for the coefficients of the boundary action $S_{bd}$ \eqref{2.5}, in the flat limit the tensors appearing in the action \eqref{A.72} must reduce to constant matrices, and, in particular, $c^{ij}$ to a constant symmetric matrix. The compatibility of the 2D Lagrangian in \eqref{A.72} with the canonical boundary structure is ensured if the relation
	\begin{equation}\label{A.73}
	\frac{\partial\mathcal L_{2D}}{\partial \dot q}=p\ ,
	\end{equation}
holds for both $q_1,p_1$ in \eqref{A.64} and $q_2,p_2$ in \eqref{CV2}, being equivalent descriptions of the boundary DOF. From \eqref{A.72} we have
	\begin{empheq}{align}\label{}
	\frac{\partial\mathcal L_{2D}}{\partial \dot \varphi}&=\sqrt{-\gamma}\left(2a^{0i}\partial_i\varphi+c^{0i}\partial_i\psi+
d^0\right)\\
	\color{black}\frac{\partial\mathcal L_{2D}}{\partial \dot \psi}&\color{black}=\sqrt{-\gamma}\left(2b^{0i}\partial_i\psi+c^{i0}\partial_i\varphi
+f^0\right)\normalcolor\ ,
	\end{empheq}
and  the request \eqref{A.73} is fulfilled if
	\begin{equation}\label{A.75}
	a^{0i}=\textcolor{black}{b^{0i}=}c^{00}=d^0\textcolor{black}{=f^0}=0\quad;\quad c^{02}\textcolor{black}{=c^{20}}=
		\hat\kappa\quad;\quad (a^{22}
	,\ b^{22},\ c^{22}
	\ \mbox{free})\ ,
	\end{equation}
where $\hat\kappa(\gamma)$ has been defined in \eqref{B.19} and $d^2,\ f^2$ are constrained by \eqref{d,f}. Eq. \eqref{A.75} represents a constraint on the metric dependence of the coefficients \eqref{coeff-a}-\eqref{coeff-c}, for which we have
\be
a^{0i} =\hat a^{0i}+\hat a\,\gamma^{0i}=0\ ;\ b^{0i}=\hat b^{0i}+\hat b\,\gamma^{0i}=0
\ ;\
c^{00}=\hat c^{00}+\hat c\,\gamma^{00}=0\ ;\
c^{02}=\hat c^{02}+\hat c\,\gamma^{02}=\hat\kappa\ . \label{A.78}
\ee
Since we do not want to impose unnecessary conditions on the induced metric $\gamma^{ij}$ (we are interested in determining if and how the 2D theory keeps memory of the bulk through the induced metric on the boundary), we must ask
	\begin{equation}\label{A.79}
	\hat a=\hat b=\hat c=0\quad\Rightarrow\quad a^{ij}=\hat a^{ij}(\gamma)\quad;\quad b^{ij}=\hat b^{ij}(\gamma)\quad;\quad c^{ij}=\hat c^{ij}(\gamma)\ .
	\end{equation}
From \eqref{A.75} and \eqref{A.79}, we also get
\begin{equation}
\hat a^{0i}=\hat b^{0i}=\hat c^{00}=0\quad;\quad\hat c^{02}=\hat c^{20}=\hat\kappa\ .
\end{equation}
Applying \eqref{A.75} and \eqref{A.79} to the action $S_{2D}$ \eqref{A.72}, we obtain
	\begin{equation}\label{A.80}
		\begin{split}
		S_{2D}[\varphi,\psi]=\int d^2X\sqrt{-\gamma}&\left[\hat a^{22}(\partial_\theta\varphi)^2+\hat b^{22}(\partial_\theta\psi)^2+2\hat\kappa\partial_t\varphi\partial_\theta\psi+\hat c^{22}\partial_\theta\varphi\partial_\theta\psi+d^2\partial_\theta\varphi+ f^2\partial_\theta\psi\right]\ ,
		\end{split}
	\end{equation}	
where all the coefficients (but $d^2$ and $f^2$) may depend on the determinant $\gamma(X)$ of the induced metric $\gamma_{ij}(X)$, but not on its components. 
The EOM of the action $S_{2D}[\varphi,\psi]$ are
	\begin{empheq}{align}
	\frac{\delta S_{2D}[\varphi,\psi]}{\delta\varphi}&=-\frac{1}{\sqrt{-\gamma}}\partial_\theta\left[\sqrt{-\gamma}\left(2\hat a^{22}\partial_\theta\varphi+2\hat \kappa\partial_t\psi+\hat c^{22}\partial_\theta\psi+
d^2\right)\right]=0\label{eom1NC}\\
	\frac{\delta S_{2D}[\varphi,\psi]}{\delta\psi}&=-\frac{1}{\sqrt{-\gamma}}\partial_\theta\left[\sqrt{-\gamma}\left(2\hat b^{22}\partial_\theta\psi+2\hat \kappa\partial_t\varphi+\hat c^{22}\partial_\theta\varphi
	+f^2\right)\right]=0\ ,\label{eom2NC}
	\end{empheq}
where we used \eqref{d,f} and the fact that  $\sqrt{-\gamma}\hat\kappa$ is constant, in order to write these equations as $\frac{1}{\sqrt{-\gamma}}\partial_\theta[...]$, $i.e.$ as a $\theta$-derivative.

\subsection{Holographic contact}\label{secHC}

We now consider the generic solutions of the BC \eqref{bcs2'}-\eqref{bcs6'}, where the bulk gauge fields $A_i(X)$ and $B_i(X)$ are now replaced by their boundary values $\partial_i\varphi(X)$ and $\partial_i\psi(X)$, defined in \eqref{2.28}-\eqref{2.29} and with $C=\hat C=0$ \eqref{C=0}-\eqref{hatC=0} :
	\begin{empheq}{align}
	\partial_\theta\psi&=-l_1\partial_t\psi\label{bcs2}\\
	\partial_\theta\varphi&=-l_2\partial_t\psi\label{bcs4}\\
	\partial_t\varphi&=-l_3\partial_t\psi\label{bcs6}\ .
	\end{empheq}
Eq.\eqref{bcs2} describes a chiral boson $\psi(X)$ moving at the 2D edge of the bulk with velocity  $v_\psi=\frac{1}{l_1}$. In the same way, by using \eqref{bcs4} in \eqref{bcs6}, we find that $\varphi(X)$ is a chiral boson as well, satisfying
	\begin{equation}\label{bcs8}
	\partial_t\varphi-\frac{l_3}{l_2}\partial_\theta\varphi=0\ ,
	\end{equation}
moving with velocity $v_\varphi=-\frac{l_3}{l_2}$. What is important to remark now is that, differently from what happens in flat spacetime \cite{Amoretti:2013xya,Maggiore:2018bxr,Maggiore:2019wie} and in analogy to the case of CS theory in curved spacetime with radial boundary \cite{Bertolini:2021iku}, on the edge of a generic bulk manifold we find two chiral bosons moving with $local$, rather than constant, velocities. In fact, both velocities, at this stage, can depend on the determinant and/or on the components of $\gamma_{ij}$ which, in general, are local quantities.
To establish the holographic contact, we consider the EOM \eqref{eom1NC}, \eqref{eom2NC} with $d^2=f^2=0$ (since the BC are homogeneously linear in the derivatives)
	\begin{empheq}{align}
	&\partial_\theta\left(2\frac{\hat a^{22}}{\hat\kappa}\partial_\theta\varphi+2\partial_t\psi+\frac{\hat c^{22}}{\hat\kappa}\partial_\theta\psi\right)=0\label{eom1NC'}\\
	&\partial_\theta\left(2\frac{\hat b^{22}}{\hat\kappa}\partial_\theta\psi+2\partial_t\varphi+\frac{\hat c^{22}}{\hat\kappa}\partial_\theta\varphi\right)=0\ .\label{eom2NC'}
	\end{empheq}
The holographic contact is realized by inserting into the EOM \eqref{eom1NC'}, \eqref{eom2NC'} the BC solutions \eqref{bcs2}-\eqref{bcs6}, which gives
\begin{empheq}{align}
-2\frac{l_2}{\hat\kappa}\hat a^{22}+2-\frac{l_1}{\hat\kappa}\hat c^{22}&=0\\
-2\frac{l_1}{\hat\kappa}\hat b^{22}-2l_3-\frac{l_2}{\hat\kappa}\hat c^{22}&=0\ .
\end{empheq}
We can write two of the three boundary parameters ($e.g.\ \hat a^{22},\hat b^{22}$) in terms of the remaining one ($\hat c^{22}$) and of the bulk coefficients ($\hat\kappa,l_i$):
	\begin{empheq}{align}
	\hat a^{22}&=+\hat\kappa\frac{1}{l_2}\left(1-\frac{l_1}{2\hat\kappa}\hat c^{22}\right)\label{aHC}\\
	\hat b^{22}&=-\hat\kappa\frac{l_3}{l_1}\left(1+\frac{l_2}{2\hat\kappa l_3}\hat c^{22}\right)\ .\label{bHC}
	\end{empheq}
Remember that $\hat{a}^{22}$ and $\hat{b}^{22}$, defined in \eqref{coeff-a} and \eqref{coeff-b}, must depend on the determinant of the induced metric only, and not on its components. On the other hand, the coefficients $l_i$ \eqref{l1}-\eqref{l3} may depend on both the determinant and the components of $\gamma_{ij}$. Therefore, we have to tune the parameters of the boundary action \eqref{2.5} in order that $\hat{a}^{22}$ and $\hat{b}^{22}$ have the right dependence. For instance, one easy way to achieve this is to set $\hat\alpha=\hat\beta=\hat\zeta=0$ in \eqref{coeff1}-\eqref{coeff3}.
These two equations are consequence of the bulk (BC)-boundary (EOM) correspondence, from which we can find out the physics of the 2D induced theory. We do this by inserting them back into the 2D action \eqref{A.80} (with $d^2=f^2=0$):
	\begin{equation}\label{S2DHC}
		S_{2D}[\varphi,\psi]=
		\kappa\int d^2X\tilde\epsilon^{012}\left\{\left[\tfrac{1}{l_2}(\partial_\theta\varphi)^2- \tfrac{l_3}{l_1}(\partial_\theta\psi)^2\right]-\tfrac{\hat c^{22}}{2\hat\kappa}\tfrac{l_1}{l_2}\left(\partial_\theta\varphi-\tfrac{l_2}{l_1}\partial_\theta\psi\right)^2+2\partial_t\varphi\partial_\theta\psi\right\}\ .
	\end{equation}

\subsection{Physical interpretation}\label{PhysInt}

The holographic contact has been imposed by inserting into the EOM the BC which represent two chiral bosons. Therefore we expect that the 2D theory should describe two chiral bosons as well. This fact appears evident by considering the following linear combination
	\begin{equation}\label{phi+-} 
	\Phi^\pm\equiv\varphi\pm \psi\ .
	\end{equation}
In terms of these new fields the action $S_{2D}[\varphi,\psi]$ \eqref{S2DHC} writes
\be
S_{2D}[\Phi^+,\Phi^-]=S_{2D}[\Phi^+]+S_{2D}[\Phi^-]+
\kappa\int d^2X\;\tilde\epsilon^{012}\partial_\theta\Phi^+\partial_\theta\Phi^-
\left(
\frac{2\hat\kappa(l_1+l_2l_3)-\hat{c}^{22}(l_1^2-l_2^2)}{4\hat\kappa l_1l_2}
\right)\ ,
\label{S+-tot}\ee
where
\be
S_{2D}[\Phi^\pm]=\frac{\kappa}{2}\int d^2X\; \tilde\epsilon^{012}
\partial_\theta\Phi^\pm(\pm\partial_t\Phi^\pm + v_\pm\partial_\theta\Phi^\pm)\ ,
\label{Lutt+-}\ee
and
\be
v_\pm=\tfrac{2\hat\kappa(l_1-l_2l_3)-\hat{c}^{22}(l_1\mp l_2)^2}{4\hat\kappa l_1l_2}\ .
\label{vpm}\ee
The action $S_{2D}[\Phi^+,\Phi^-]$ \eqref{S+-tot} decouples into the sum of the  Luttinger actions \eqref{Lutt+-}
\be
S_{2D}[\Phi^+,\Phi^-]=S_{2D}[\Phi^+]+S_{2D}[\Phi^-]\ ,
\label{S2D}\ee
provided that the following condition on the parameters of the theory holds
	\begin{equation}\label{decoup}
2\hat\kappa(l_1+l_2l_3)-\hat{c}^{22}(l_1^2-l_2^2)=0\ .
	\end{equation}
Once decoupled, we may identify the fields $\Phi^+(X)$ and $\Phi^-(X)$ as Right (R) and Left (L) modes moving at the radial edge of the 3D bulk theory with velocities $\pm v_\pm$ respectively \cite{Kane}, where
\be
v_\pm=\frac{1\mp l_3}{l_2\pm l_1}\qquad\mbox{if}\  l_1^2-l_2^2\neq 0\ ,
\label{}\ee
and
\bea
v_+=\frac{1}{l_1}\ \ ;\ \ v_-=\frac{1}{l_1}-\frac{\hat{c}^{22}}{\hat\kappa}&\qquad\mbox{if}\  & l_1=l_2\ \ ,\ \ l_3=-1\label{}\\
v_+=-\frac{1}{l_1}+\frac{\hat{c}^{22}}{\hat\kappa}\ \ ;\ \ v_-=-\frac{1}{l_1}\ \ \ \ &\qquad\mbox{if}\  & l_1=-l_2\ \ ,\ \ l_3=1\ . \label{}
\eea
As a consequence of the holographic contact, the metric dependence of the boundary parameters  \eqref{A.79} is transferred to the $l_i$ coefficients through \eqref{aHC} and \eqref{bHC}. This makes $v_\pm$ depend on the determinant of the induced metric $v_\pm=v_\pm(\gamma)$. We therefore remark the crucial point that the fact of dealing with a curved bulk spacetime has the primary consequence that the velocities of the edge modes depend on both time and space $v_\pm=v_\pm(t,\theta)$, differently to what happens for flat backgrounds.
Hence, from $v_\pm$ we see that the edge action $S_{2D}$ \eqref{S2D} may describe three classes of physical situations, tuned by the $local$ bulk parameters $l_i$ and by $\hat{c}^{22}$~:
\begin{enumerate}
\item $\pmb{v_+v_->0}$: LR movers with opposite velocities.\\
It is realized if 
\be
\frac{1-l^2_3}{l_2^2-l_1^2}>0\qquad\mbox{if}\  l_2^2-l_1^2\neq 0\ ,
\label{}\ee
or 
\be
\hat\kappa-\hat{c}^{22}l_1>0\qquad\mbox{if}\   l_1=\pm l_2\ \ ,\ \ l_3=\mp 1\ .
\label{}\ee
This situation describes generic chiral Luttinger liquids \cite{Wen:2004ym}, but also helical ones \cite{Wu2006HelicalLA}.
In fact
ordinary TI \cite{moorenature,Hasan:2010xy,Hasan:2010hm,Qi-Zhang,Cho:2010rk}, characterized by edge modes moving in opposite directions with equal velocities
\be
v_+=v_-\qquad\mbox{(Topological Insulators)}\ ,
\label{v+=v-}\ee
fall into this category. It is easy to see that the condition \eqref{v+=v-} is satisfied provided that
\bea
l_1+l_2l_3&=&0 \label{l1+l2l3=0}\\
\hat{c}^{22}&=&0\ .
\label{c22=0}
\eea
The equal and opposite edge velocities therefore are
\be
v_+=v_-=\frac{1}{l_2}\ ,
\label{}\ee
which still, for a generic bulk manifold, may have a spacetime dependence.
\item $\pmb{v_+v_-<0}$: LR movers in the same direction.\\
It is realized if 
\be
\frac{1-l^2_3}{l_2^2-l_1^2}<0\qquad\mbox{if}\  l_2^2-l_1^2\neq 0
\label{3.59}\ee
or 
\be
\hat\kappa-\hat{c}^{22}l_1<0\qquad\mbox{if}\   l_1=\pm l_2\ \ ,\ \ l_3=\mp 1\ .
\label{3.60}\ee
Also in this case we can recover the particular case of a pair of Hall systems \cite{wen2}, with edge excitations moving in the same direction with the same velocity
\be
v_+=-v_-\qquad\mbox{(pair of Hall systems)}\ ,
\label{v+=-v-}\ee
realized if
\bea
l_2+l_1l_3 &=& 0 \label{}\\
\hat{c}^{22}&=& \frac{2\hat\kappa}{l_1}\ . \label{}
\eea
The velocities of the edge modes in this case are
\be
v_+=-v_-=\frac{1}{l_1}\ .
\label{}\ee
\item $\pmb{v_+v_-=0}$: L or R mover not moving, which characterizes the Quantum Anomalous Hall (QAH) Insulators  \cite{Qi-Zhang}. This happens when
\be
l_3=\pm1\qquad\mbox{if}\  l_1\mp l_2\neq0\ ,
\label{}\ee
which means
\be
v_\pm=0\ ;\  
v_\mp=\frac{2}{l_2\mp l_1}\ ;
\ \hat{c}^{22}=\frac{2\hat\kappa}{l_1\mp l_2}\ .
\label{v+-=0}\ee
\end{enumerate}
Some comments are in order. First of all we notice that, since the BF coupling constant $\kappa$ must be positive,
 $\hat{c}^{22}=0$ uniquely identifies L and R modes moving on the edge of the 3D bulk with opposite velocities ($\pmb{v_+=v_-}$). 
Hence, as for chiral velocities, $\hat{c}^{22}$ should be determined either by a phenomenological input or by a symmetry principle. Now, TI belong to this class of edge excitations, and are topological phases of electrons which respect Time Reversal ($T$) symmetry \cite{Qi-Zhang,Cho:2010rk}. This suggests that $\hat{c}^{22}=0$ might be related to the conservation of $T$-symmetry and, conversely, $\hat{c}^{22}\neq 0$ to its violation, like it happens, for instance, in the case of the QAH Insulators \cite{Qi-Zhang}, described by case 3.
We shall come back to this point in the next Section. 
A second comment comes from the fact that 
it is necessary that the Hamiltonian corresponding to the action \eqref{S2D} is positive definite. This request yields constraints on the bulk parameters of the model, $i.e.$ the ``coupling'' constant $\hat\kappa$ \eqref{B.19}, the parameters $\alpha^{ij}$ \eqref{coeff1}, $\beta^{ij}$ \eqref{coeff2} and $\zeta^{ij}$ \eqref{coeff3} appearing in $S_{bd}$ \eqref{2.5}, together with the parameter $\hat{c}^{22}$. We recall that the parameters $l_i$ depend on the bulk through \eqref{l1}, \eqref{l2} and \eqref{l3} and appear in the action $S_{2D}[\Phi^+,\Phi^-]$ through $v_\pm$.
The canonical variables defined in \eqref{A.64} and \eqref{CV2} in terms of $\Phi^\pm$ \eqref{phi+-} write
\begin{equation}
		\begin{split}
		q_1&=\frac{\Phi^++\Phi^-}{2}\quad;\quad p_1=\frac{\kappa\tilde\epsilon^{012}}{2}\partial_\theta\left(\Phi^+-\Phi^-\right)\\
		q_2&=\frac{\Phi^+-\Phi^-}{2}\quad;\quad p_2=\frac{\kappa\tilde\epsilon^{012}}{2}\partial_\theta\left(\Phi^++\Phi^-\right)\ .
		\end{split}
	\end{equation}
The Hamiltonian density of the model therefore is
	\begin{equation}\label{T00}
		\begin{split}
		\mathcal H_{2D}&=p_1\dot q_1+p_2\dot q_2-\mathcal L_{2D}\\
		&=-\frac{1}{2\kappa\tilde\epsilon^{012}}\left[v_+(p_1+p_2)^2+v_-(p_1-p_2)^2\right]\\
		&=-\frac{\kappa\tilde\epsilon^{012}}{2}\left[v_+(\partial_\theta\Phi^+)^2+v_-(\partial_\theta\Phi^-)^2\right]\ .
		\end{split}
	\end{equation}
Positive energy density means $\mathcal H_{2D}>0$,  and since $\kappa>0$ \eqref{2.67}, requiring the coefficients of the squared terms to be positive gives the following constraint
	\begin{equation}
	\mathcal H_{2D}>0\quad\Leftrightarrow\quad v_+\leq0,\ v_-\leq0\ .\label{positiveH}
	\end{equation}
Therefore, we observe that the physical situation of edge modes moving in the same direction ($\pmb{v_+v_-<0}$) is not compatible with the positivity conditions \eqref{positiveH}. The fact that the Hamiltonian is not bounded by below would lead us to discard this case, leaving us only with the cases 1 and 3. 

\section{The role of Time-Reversal symmetry}\label{secTRBC}

In this Section we present two alternative ways of introducing the Time Reversal $T$-symmetry in the theory with boundary. It will turn out that these two approaches, although seemingly quite different, are indeed equivalent. $T$-transformation is defined in the usual way as $Tx^0=-x^0$. Due to the invariance of the line element $ds^2$, we have
	\begin{equation}\label{Tgamma}
	T\gamma_{t\theta}=-\gamma_{t\theta}\footnote{Notice that if the metric is stationary, $i.e.\ \partial_t\gamma_{ij}=0$, then the $T$-invariance \eqref{Tgamma} requires that $\gamma_{t\theta}=0$. }\ .
	\end{equation}
The only components of the gauge fields which change sign under $T$ are:
	\begin{equation}
TA_r(t,r,\theta)=-A_r(-t,r,\theta)\ \ ;\ \ 
TA_\theta(t,r,\theta)=-A_\theta(-t,r,\theta)\ \ ;\ \ 
TB_t(t,r,\theta)=-B_t(-t,r,\theta)\ .\label{Tgf}
	\end{equation}
According to this definition $A_\mu(x)$ may be associated to an electric potential and $B_\mu(x)$ to a spin current \cite{Amoretti:2014iza}. 

\subsection{$TS_{bd}=S_{bd}$}

It is immediate to see that the bulk action  $S_{BF}$ \eqref{2.1} is $T$-invariant\footnote{Other choices of $T$ are possible which leave $S_{BF}$ invariant, like for instance $TA_t=-A_t,\ TB_r=-B_r,\ TB_\theta=-B_\theta$, which correspond to $A_\mu\leftrightarrow B_\mu$. One can show that these choices are equivalent \cite{Amoretti:2014iza}.}, and it is interesting to study which are the consequences of imposing Time-Reversal also on the boundary term  $S_{bd}$ \eqref{2.5}, $i.e.$ 
	\begin{equation}\label{Tsbd inv}
	TS_{bd}=S_{bd}\ .
	\end{equation}
Due to \eqref{Tgamma}, to the fact that $T\gamma=\gamma$ and to the form of the coefficients \eqref{coeff1}, \eqref{coeff2} and \eqref{coeff3}, the parameters appearing in $S_{bd}$ \eqref{2.5} which transform non-trivially under $T$ are
	\begin{empheq}{align}
	T\alpha^{02}=&\hat\alpha^{02}-\hat\alpha\gamma^{02}\label{xTAcoeff1}\\
	T\beta^{02}=&\hat\beta^{02}-\hat\beta\gamma^{02}\label{xTAcoeff2}\\
	T\zeta^{02}=&\hat\zeta^{02}-\hat\zeta\gamma^{02}\label{xTAcoeff3}\\
	T\zeta^{20}=&\hat\zeta^{20}-\hat\zeta\gamma^{20}\ .\label{xTAcoeff4}
	\end{empheq}
Requiring the invariance \eqref{Tsbd inv}, from \eqref{Tgf} and \eqref{xTAcoeff1}-\eqref{xTAcoeff4}, we get the following constraints
	\begin{equation}\label{Thp}
	\hat\alpha^{02}=\hat\alpha^{20}=0\quad;\quad\hat\beta^{02}=\hat\beta^{20}=0\quad;\quad\zeta^{00}=0\quad;\quad\zeta^{22}=0\quad;\quad\hat\zeta=0\ .
	\end{equation}
The resulting $T$-invariant boundary term $S_{bd}$ \eqref{2.5} is
	\begin{equation}\label{xB.23'}
		\begin{split}
		S_{bd}=\int d^3x\,\sqrt{-\gamma}\,\delta(r-R)&\left[\frac{\gamma^{ij}}{2}\left(\hat\alpha A_i A_j+\hat\beta B_iB_j\right)+\frac{\hat\alpha^{00}}{2}A_t A_t+\frac{\hat\alpha^{22}}{2}A_\theta A_\theta+\right.\\
		&\left.\ +\frac{\hat\beta^{00}}{2}B_tB_t+\frac{\hat\beta^{22}}{2}B_\theta B_\theta+\hat\zeta^{02}A_t B_\theta+\hat\zeta^{20}A_\theta B_t\right]\ ,
		\end{split}
	\end{equation}
and we recall that all the coefficients appearing in \eqref{xB.23'} might still depend on the determinant $\gamma$ of the induced metric $\gamma_{ij}$, being therefore local quantities and not simply constants. It will be interesting to investigate which are the consequences, if any, of imposing $T$ on $S_{bd}$ on the holographically induced 2D theory.

\subsubsection{Generic non-diagonal metric ${\gamma_{t\theta}\neq0}$}\label{secTnondiag}

 As a consequence of \eqref{Thp}, in the hypothesis of $\gamma_{t\theta}\neq0$ and if $\hat\alpha$ and/or $\hat\beta$ are/is non-vanishing, the BC \eqref{BC1} and \eqref{BC2} become
	\begin{empheq}{align}
	\left.\alpha^{00}A_t+\hat\alpha\gamma^{t\theta}A_\theta+(\hat\zeta^{02}-\hat\kappa)B_\theta\right|_{r=R}=&0\label{BCT1}\\
	\left.\hat\alpha\gamma^{t\theta}A_t+\alpha^{22}A_\theta+(\hat\zeta^{20}+\hat\kappa)B_t\right|_{r=R}=&0\label{BCT2}\\
	\left.\hat\zeta^{20}A_\theta+\beta^{00}B_t+\hat\beta\gamma^{t\theta}B_\theta\right|_{ r=R}=&0\label{BCT3}\\
	\left. \hat\zeta^{02}A_t+\hat\beta\gamma^{t\theta}B_t+\beta^{22}B_\theta\right|_{ r=R}=&0\ ,\label{BCT4}
	\end{empheq}
or, using the notation already adopted in \eqref{B.16},
	\begin{equation}\label{BCT5}
	\left.\Lambda_T^{IJ}X_J\right|_{r=R}=0\ ,
	\end{equation}
where $\Lambda_T$ is the matrix \eqref{Lambda} evaluated at \eqref{Thp}. Notice the explicit dependence on the off-diagonal component $\gamma_{t\theta}$ of the induced metric and on $\hat\alpha$ or $\hat\beta$. The linear system \eqref{BCT5} has nontrivial solutions if $\det\Lambda_T=0$, $i.e.$
\begin{equation}
\begin{split}
0=\det\Lambda_T=&\hat\alpha\left(\gamma^{t\theta}\right)^2\left\{-\hat\alpha\det\beta+\hat\beta\left[ \hat\zeta^{20} \hat\kappa -  \hat\zeta^{02} \left(2   \hat\zeta^{20}+\hat\kappa \right)\right]\right\}+ \\
&+ \alpha^{00} \left[ \alpha^{22} \det\beta-  \beta^{22}   \hat\zeta^{20} \left(  \hat\zeta^{20}+\hat\kappa \right)\right]+  \hat\zeta^{02}\left(  \hat\zeta^{02}-\hat\kappa \right) \left[  \hat\zeta^{20} (  \hat\zeta^{20}+\hat\kappa )-  \alpha^{22}   \beta^{00}\right]\ .
\end{split}
\end{equation}
It is easily seen that the solutions \eqref{bcs2'}-\eqref{bcs6'} are recovered, with
	\begin{equation}\label{vi'}
	l_1\to l_1|_{\eqref{Thp}}\quad;\quad l_2\to l_2|_{\eqref{Thp}}\quad;\quad l_3\to l_3|_{\eqref{Thp}}\ ,
	\end{equation}
and $l_{1,3}|_{\eqref{Thp}}\propto\gamma^{t\theta}$, as shown in Appendix \ref{appl}. Notice that for a diagonal (but not necessarily static) metric ($\gamma_{t\theta}=0$) or when $\hat\alpha=\hat\beta=0$, these solutions must be discarded, since they imply Dirichelet BC on both fields ($l_{1,3}|_{\eqref{Thp}}=0\Rightarrow B_\theta(X)=A_t(X)=0$), which, as we already remarked, would trivialize the 2D physics. The case of diagonal metric or $\hat\alpha=\hat\beta=0$ will be analyzed in the next Subsection. The procedure we followed to recover the 2D theory does not change: the bulk-boundary correspondences \eqref{aHC} and \eqref{bHC} still hold, with the replacements \eqref{vi'}. 
From \eqref{l1'} and \eqref{l3'} we see that, due to the $T$-invariance request \eqref{Thp}, the coefficients $l_1$ and $l_3$ appearing in the BC \eqref{bcs2'} and \eqref{bcs6'}
explicitly depend on $\gamma^{t\theta}$. As a consequence, $\hat{a}^{22}$ \eqref{aHC} and $\hat{b}^{22}$ \eqref{bHC} would depend on $\gamma^{t\theta}$ as well, but we know that, due to the holographic contact, the coefficients appearing in the action $S_{2D}$ \eqref{A.80} should depend on the determinant of the induced metric only, and not on its components. The only way to realize this is to set $\hat{c}^{22}=0$. In fact, in this case we have
\be
\hat{c}^{22}=0 \quad\Rightarrow \quad\frac{l_1|_{\eqref{Thp}}}{l_3|_{\eqref{Thp}}}=-l_2|_{\eqref{Thp}}\ ,
\label{c22 l1+l2l3}\ee
which does not depend on $\gamma^{t\theta}$, and, from \eqref{aHC} and \eqref{bHC}, 
	\begin{equation}\label{HCT}
	\hat a^{22}=\frac{1}{l_2|_{\eqref{Thp}}}\hat\kappa\quad;\quad\hat b^{22}=-\frac{l_3|_{\eqref{Thp}}}{l_1|_{\eqref{Thp}}}\hat\kappa\quad \ .
	\end{equation}
Eq. \eqref{c22 l1+l2l3} coincides with eqs \eqref{c22=0} and \eqref{l1+l2l3=0}, which belong to the case 1 considered in the previous Section, where we have seen that the physical situation described by the decoupled action $S_{2D}[\Phi^+,\Phi^-]$ \eqref{S2D} together with the conditions \eqref{c22=0} and \eqref{l1+l2l3=0}, is that of
a Luttinger model for two chiral currents with non-constant and opposite velocities. We thus established a link between the parameter  $\hat c^{22}$ and $T$-invariance on the boundary, which enforces the physical interpretation as edge states of TI, as anticipated in the previous Section. 

\subsubsection{Diagonal metric $\gamma_{t\theta}=0$}\label{secTdiag}

When considering a diagonal metric or boundary coefficients \eqref{coeff1}-\eqref{coeff3} which depend at most on the determinant of the metric ($i.e.$ when $\gamma_{t\theta}=0$ or $\hat\alpha=\hat\beta=0$), the BC \eqref{BCT1}-\eqref{BCT4} become
	\begin{empheq}{align}
	\left.\alpha^{00}A_t+(\hat\zeta^{02}-\hat\kappa)B_\theta\right|_{r=R}=&0\label{BCTs1}\\
	\left.\alpha^{22}A_\theta+(\hat\zeta^{20}+\hat\kappa)B_t\right|_{r=R}=&0\label{BCTs2}\\
	\left.\hat\zeta^{20}A_\theta+\beta^{00}B_t\right|_{ r=R}=&0\label{BCTs3}\\
	\left. \hat\zeta^{02}A_t+\beta^{22}B_\theta\right|_{ r=R}=&0\ ,\label{BCTs4}
	\end{empheq}
which represent two systems of homogeneous linear equations. 
Non-Dirichelet solutions are
	\begin{empheq}{align}
	A_t(X)=&-l_aB_\theta(X)\label{xBCT1}\\
	A_\theta(X)=&-l_bB_t(X)\ ,\label{xBCT2}
	\end{empheq}
where
\be
l_a=\frac{\beta^{22}}{\hat\zeta^{02}}\ \ \mbox{and}\ \ l_b=\frac{\beta^{00}}{\hat\zeta^{20}}\ ,
\label{}\ee
provided that
\bea
\left(\hat\zeta^{02}-\hat\kappa\right)\hat\zeta^{02}-\alpha^{00}\beta^{22} &=& 0 \label{}\\
\left(\hat\kappa+\hat\zeta^{20}\right)\hat\zeta^{20}-\alpha^{22}\beta^{00} &=&0\ .\label{}
\eea
Following the same steps described in Section 3, we still land on the action $S_{2D}$  \eqref{A.80}. The holographic contact is realized crossing the EOM \eqref{eom1NC} and \eqref{eom2NC} with the BC \eqref{xBCT1} and \eqref{xBCT2}, which in terms of the boundary fields $\varphi(X)$, $\psi(X)$ read:
	\begin{empheq}{align}
	\partial_t\varphi=&-l_a\partial_\theta\psi\label{bcsT1}\\
	\partial_\theta\varphi=&-l_b\partial_t\psi\ .\label{bcsT2}
	\end{empheq}
The correspondence is achieved if
	\begin{equation}\label{HClambT'}
	\hat a^{22}=\frac{1}{l_b}\hat\kappa\quad;\quad\hat b^{22}=l_a\hat\kappa\quad;\quad\hat c^{22}=0\ .
	\end{equation}
From the properties of the tensors $\hat a^{ij},\hat b^{ij},\hat c^{ij}$ in \eqref{A.79}, it is immediate to check that $l_{a,b}$ depend only on the determinant of the induced metric $\gamma$, $i.e.\ l_{a,b}=l_{a,b}(\gamma)$. The 2D action \eqref{A.80} decouples into a pair of Luttinger models, provided that the following condition holds
\begin{equation}
l_a=\frac{1}{l_b}\ ,
\end{equation}
and describes two chiral modes travelling on the edge of the 3D bulk with equal and opposite local velocities $v_\pm=l_a(\gamma)$
	\begin{empheq}{align}
	\partial_t\Phi^++l_a\partial_\theta\Phi^+&=0\label{bcsT1+}\\
	\partial_t\Phi^--l_a\partial_\theta\Phi^-&=0\ ,\label{bcsT2-}
	\end{empheq}
characterizing TI \cite{Hasan:2010xy,Qi-Zhang,Cho:2010rk}. \\

\subsection{Inherited $T$-transformation}\label{T2D}

It is possible to impose the $T$-symmetry on the theory with boundary in an alternative way with respect to what we did in the previous Subsection, reaching the same physical conclusions ($i.e.$ chiral edge modes moving with opposite velocities as the unique physical outcome of putting a boundary onto 3D BF theory). The bulk gauge fields $A_\mu(x)$ and $B_\mu(x)$ transform under $T$ according to \eqref{Tgf}. Consequently, the boundary scalar fields $\psi(X)$ and $\varphi(X)$, being defined by \eqref{2.28} and \eqref{2.29}, should inherit the following $T$-transformations
\begin{equation}\label{Tphi}
	T\varphi(t,\theta)=-\varphi(-t,\theta)\quad;\quad T\psi(t,\theta)=\psi(-t,\theta)\ ,
\end{equation}
hence, due to \eqref{phi+-}, we have
\begin{equation}\label{Tphi+-}
	T\Phi^+(t,\theta)=-\Phi^-(-t,\theta)\quad;\quad T\Phi^-(t,\theta)=-\Phi^+(-t,\theta)\ .
\end{equation}
The action $S_{2D}[\Phi^+,\Phi^-]$ \eqref{S+-tot} is $T$-invariant if
\be
v_+=v_-\ ,
\label{}\ee
and if the decoupling condition \eqref{decoup} holds. This is precisely the situation considered in case 1 treated in Section 3.4, uniquely identified by the vanishing of the parameter $\hat{c}^{22}$ \eqref{c22=0}. \\

We therefore checked our guess on the peculiar role played by the parameter $\hat{c}^{22}$ appearing in the action $S_{2D}[\varphi,\psi]$ \eqref{A.80}, and also that the two alternative ways described in this Section of imposing the $T$-symmetry (on the boundary term $S_{bd}$ \eqref{2.5} or directly on $S_{2D}[\varphi,\psi]$ through the defining relations \eqref{2.28} and \eqref{2.29}) are indeed physically equivalent. 
Lastly, we remark that the physical situation represented by the TI, $i.e.$ the existence of chiral edge modes moving with equal and opposite velocities, is singled out 
by imposing $T$-symmetry (in either way), while asking that the Hamiltonian ${\cal H}_{2D}$ is bounded by below, as discussed in Section 3.4, admits a larger class of physical situations, including $T$-breaking effects, but only for opposite-moving modes. The possibility of edge modes moving in the same direction is ruled out by both argumentations, and $T$-symmetry is more restrictive than the positive energy condition. According to this analysis, case 3 is not compatible with $T$-symmetry, however lower-bounded Hamiltonian does admit the possibility of a single-moving edge state. Hence it can be interpreted as a $T$-breaking effect associated to QAH insulators \cite{Qi-Zhang}. In the same way the situations with $\hat c^{22}\neq0$ belonging to case 1 can be seen as other examples of symmetry breaking effects in Quantum Spin Hall systems \cite{Hasan:2010xy,Qi-Zhang,Wu2006HelicalLA}.\\

It is known \cite{Qi-Zhang,Cho:2010rk} that $T$-symmetry is peculiar to the  edge states of TI in 2D, which are described by a helical Luttinger  model \cite{Wu2006HelicalLA}. This is exactly the situation we observed for $\hat c^{22}=0$ \eqref{c22=0}. Therefore we may finally claim that BF model with boundary together with $T$-invariance is an effective description of the edge states of TI with possibly non-constant chiral velocities. The existence of edge accelerated modes is a direct consequence of the bulk/boundary correspondence in curved spacetime: flat spacetime analysis provides only for constant velocities. The $\hat c^{22}\neq0$ situation now can be related to $T$-breaking phenomena in helical Luttinger liquids \cite{Wu2006HelicalLA}. For instance (as we already remarked) we can associate the case 3 of Section \ref{PhysInt} to the Anomalous Quantum Hall Effect \cite{Qi-Zhang, Liu, Yu}.

\section{Conclusions}\label{secSumm+Disc}

In this paper we considered the abelian 3D BF model on a manifold described by a generic metric, with a radial boundary which spoils the topological character of the theory. Following a method introduced by Symanzik, we added to the action a boundary term constrained only by locality and power counting, in order to find out the most general boundary conditions on the two gauge fields involved in the theory. The boundary breaks gauge invariance, and this reflects in two broken Ward identities, from which we identified the boundary degrees of freedom, represented by two scalar fields, deriving also a semidirect sum of two Ka\c{c}-Moody algebras formed by two on-shell conserved currents. As in the Chern-Simons case, the central charge of the Ka\c{c}-Moody algebra is proportional to the inverse of the BF coupling, and this constrains the BF coupling to be positive. Once written in terms of the scalar boundary degrees of freedom, the Ka\c{c}-Moody algebra can be interpreted as the commutation relation of canonical variables, which led us to derive the corresponding 2D action. The bulk/boundary holographic correspondence is achieved by matching the boundary conditions on the 3D gauge fields and the equations of motion of the 2D action. The resulting 2D action is a complicated functional of the pair of scalar fields but, by  means of a simple linear redefinition, the action can be written in terms of two Luttinger actions for two chiral fields, plus a mixed term, which we force to vanish by imposing a decoupling condition. At this point, the parameters which survived to the holographic contact and to the decoupling condition allow for interesting physical interpretations of the theory, which may indeed describe~:
\begin{enumerate}

\item two edge excitations moving in opposite directions. This is realized in 
Hall systems, like Fractional Quantum Hall with $\nu=1-1/n$ \cite{wen2}, and edge modes of Quantum Spin Hall systems, like Topological Insulators (when $v_+=v_-$), possibly interacting \cite{Calzona}, or nanowires \cite{meng} with additional magnetic fields acting on the velocities up to switching one off \cite{Streda,Heedt}. In higher dimensions an effect of renormalization of chiral velocities ($i.e.\ v_+v_->0$) can be achieved by adding magnetic fields \cite{bernevig,goerbig2}, or by structural deformations \cite{goerbig};

\item two chiral bosons moving in the same direction. This physically corresponds to Hall systems like, for instance, Quantum Hall with $\nu=2$, or Fractional Quantum Hall with $\nu=2/5$ \cite{Bocquillon,wen2,ferraro} possibly with non-constant interactions or confining potentials \cite{Wen:1989mw,Wen:1990qp,Kane95,Hashi18,Wen:1991ty};

\item one static and one moving mode. This is the situation of 
Quantum Anomalous Hall, where magnetic impurities break the $T$-symmetry of TI \cite{Qi-Zhang, Liu, Yu}. This could also be explained as  an extremal effect of a magnetic field acting on the modes of the nanowires mentioned above.
\end{enumerate}
In all cases described above, the chiral velocities depend on the induced metric on the boundary, hence on time and space. Therefore, their local nature is a direct consequence of the non-flatness of the bulk metric. Moreover, we calculated the Hamiltonian corresponding to the 2D action and we found that the request of the existence of a lower bound rules out the case 2~: Hall systems with parallel velocities, like Fractional Quantum Hall with $\nu=2/5$ \cite{wen2} or Integer Quantum Hall with $\nu=2$ \cite{ferraro}, cannot be described by a BF theory with boundary. We then considered the role of Time Reversal symmetry, which distinguishes the two main Schwarz-type TQFT: $T$-violating Chern-Simons and $T$-respecting BF. $T$-symmetry can be introduced in two different ways: from the bulk side asking that the ``Symanzik's'' boundary term of the 3D action is $T$-invariant, which has consequences on the subsequent steps till the holographically induced 2D action. $T$-symmetry may be imposed also on the boundary side directly on the 2D action, starting from the definition of the scalar degrees of freedom. These two seemingly inequivalent ways of requiring $T$-invariance lead to the same outcome: the only physical case which respects $T$-symmetry is a subclass of case 1 above, namely the one involving two edge excitations moving in opposite directions {\it with the same velocity}, which, in the general case of non-Minkowskian bulk metric described in this paper, {\it might be local}, $i.e.$ time and space dependent. This is the case of Topological Insulators. Hence, imposing $T$-symmetry is much more restrictive than asking for a lower bounded Hamiltonian. Topological phases of matter displaying accelerated edge modes have been observed for instance in the Integer Quantum Hall with $\nu=2$ \cite{Bocquillon}. But, to our knowledge, generalized Topological Insulators with accelerated chiral edge modes have not been discovered yet. In this paper we predict that they should, and we presented a theoretical framework for their existence, as a direct consequence of a non-Minkowskian bulk background. In our opinion, this represents a cleaner alternative to adding an {\it ad hoc} local potential to the pair of the decoupled Luttinger actions, which would spoil the whole holographic construction described in this paper.

\section*{Acknowledgments}

We thank Dario Ferraro, Niccol\`o Traverso Ziani, Paolo Meda and Giandomenico Palumbo for enlightening discussions. E.B. and F.F. thank the Galileo Galilei Institute (Firenze, Italy) and the TU Wien (Austria), respectively, for hospitality during part of this work, which has been partially supported by the INFN Scientific Initiative GSS: ``Gauge Theory, Strings and Supergravity''. E.B. is supported by MIUR grant ``Dipartimenti di Eccellenza'' (100020-2018-SD-DIP-ECC\_001). F.F. acknowledges the support of the Erasmus+ programme of the European Union.

\appendix

\section{Solutions of the boundary conditions \eqref{B.16}}\label{appl}
The BC for the 3D bulk theory are given by \eqref{B.16}:
	\begin{equation}
	\left.\Lambda^{IJ}X_J\right|_{r=R}=0\ ,
	\end{equation}
where $I,J=\{i;j\}=\{0,2;0,2\}$,
\begin{equation}
\Lambda^{IJ}\equiv\left(\begin{array}{cc}
\alpha^{ij} & \zeta^{i j}-\kappa\epsilon^{i1 j} \\
\zeta^{j i} & \beta^{ i j} \\
\end{array}\right)=\left(
\begin{array}{cccc}
\alpha^{00} & \alpha^{02} & \zeta^{00} & \zeta^{02}-\hat\kappa\\
 \alpha^{20} & \alpha^{22} & \zeta^{20}+\hat\kappa & \zeta^{22}\\
 \zeta^{00} & \zeta^{20} & \beta^{00} & \beta^{02}\\
 \zeta^{02} & \zeta^{22}& \beta^{02} & \beta^{22}
\end{array}
\right)\ , 
\end{equation}
and
\begin{equation}
X_J\equiv\left(\begin{array}{c}
A_j\\
B_{j}
\end{array}\right)\ ,
\end{equation}
and $\hat\kappa$ is given by \eqref{B.19}. Being the linear system of equations \eqref{B.16} homogeneous, three of the four components $X_J$ can be written in terms of the fourth, provided that
	\begin{equation}
	\det\Lambda=0\ .
	\end{equation}
The choice we make is
	\begin{empheq}{align}
	B_\theta(X)&= -l_1B_t(X)\\
	A_\theta(X)&= -l_2B_t(X)\\
	A_t(X)&= -l_3B_t(X)\ ,
	\end{empheq}
with
\begin{empheq}{align}
{l_1}&\equiv-\frac{-\alpha ^{00} (\beta ^{00} \zeta ^{22}- \beta ^{02} \zeta ^{20})+\alpha ^{02}( \beta ^{00} \zeta ^{02}- \beta ^{02} \zeta ^{00})+\zeta ^{00} \det\zeta}{\alpha ^{00} (\beta ^{02} \zeta ^{22}-\beta ^{22} \zeta ^{20})+\alpha ^{02} (\beta ^{22} \zeta ^{00}-\beta ^{02} \zeta ^{02})-(\zeta ^{02}-\hat\kappa ) \det\zeta}\label{l1}\\
{l_2}&\equiv-\frac{\alpha ^{00}\det\beta-\beta ^{00}\zeta ^{02}(\zeta ^{02}-\hat\kappa)-\hat\kappa  \beta ^{02} \zeta ^{00}+2 \beta ^{02} \zeta ^{00} \zeta ^{02}-\beta ^{22}( \zeta ^{00})^2}{\alpha ^{00} (\beta ^{02} \zeta ^{22}-\beta ^{22} \zeta ^{20})+\alpha ^{02} (\beta ^{22} \zeta ^{00}-\beta ^{02} \zeta ^{02})-(\zeta ^{02}-\hat\kappa ) \det\zeta}\label{l2}\\
{l_3}&\equiv-\frac{-\alpha ^{02} \det\beta+( \zeta ^{02} -\hat\kappa)(  \beta ^{00} \zeta ^{22}-\beta ^{02} \zeta ^{20})+\zeta ^{00} (\beta ^{22}\zeta ^{20}-\beta ^{02}\zeta ^{22})}{\alpha ^{00} (\beta ^{02} \zeta ^{22}-\beta ^{22} \zeta ^{20})+\alpha ^{02} (\beta ^{22} \zeta ^{00}-\beta ^{02} \zeta ^{02})-(\zeta ^{02}-\hat\kappa ) \det\zeta}\ ,\label{l3}
\end{empheq}
where the request of non-Dirichelet solutions \cite{Blasi:2017pkk,Blasi:2015lrg} implies the requirements $l_i\neq0$ and $l_i^{-1}\neq0$.
We remind that these coefficients are local, depending on the induced metric determinant and/or components from the bulk parameters. In the case of $T$-invariant $S_{bd}$ \eqref{xB.23'} discussed in Section \ref{secTRBC}, the coefficients \eqref{vi'} are given by
\begin{empheq}{align}
{l_1}|_{\eqref{Thp}}&=-\gamma^{t\theta}\frac{\hat\beta  \alpha ^{00} \hat\zeta ^{20}+\hat\alpha  \beta ^{00} \hat\zeta ^{02}}{-\alpha ^{00} \beta ^{22} \hat\zeta ^{20}+\hat\zeta ^{02}\left[\hat\zeta ^{20} (\hat\zeta ^{02}-\hat\kappa )-\hat\alpha\hat\beta  \left(\gamma^{t\theta}\right)^2 \right]}\label{l1'}\\
l_2|_{\eqref{Thp}}&=-\frac{\alpha ^{00}\det\beta-\beta ^{00}\hat \zeta ^{02} (\hat\zeta ^{02}-\hat\kappa )}{-\alpha ^{00} \beta ^{22} \hat\zeta ^{20}+\hat\zeta ^{02}\left[\hat\zeta ^{20} (\hat\zeta ^{02}-\hat\kappa )-\hat\alpha\hat\beta  \left(\gamma^{t\theta}\right)^2 \right]}\label{l2'}\\
l_3|_{\eqref{Thp}}&=-\gamma^{t\theta}\frac{-\hat\alpha \det\beta-\hat\beta  \hat\zeta ^{20} (\hat\zeta ^{02}-\hat\kappa )}{-\alpha ^{00} \beta ^{22} \hat\zeta ^{20}+\hat\zeta ^{02}\left[\hat\zeta ^{20} (\hat\zeta ^{02}-\hat\kappa )-\hat\alpha\hat\beta  \left(\gamma^{t\theta}\right)^2 \right]}\ .\label{l3'}
\end{empheq}

\end{document}